\documentclass[aps,prl,twocolumn,showpacs,preprintnumbers,superscriptaddress]{revtex4}
\usepackage{bm,amsmath,amssymb,latexsym,mathrsfs,enumerate,color}
\usepackage[mathcal]{euscript}
\usepackage[hypertex]{hyperref}
\usepackage{graphicx}
\renewcommand{\vec}[1]{\bm{#1}}
\begin{document}

\title{Controlled vortex core switching in a magnetic nanodisk by a rotating field}

\author{Volodymyr P. Kravchuk}
 \affiliation{National Taras Shevchenko University of Kiev, 03127 Kiev, Ukraine}

\author{Denis D. Sheka}
 \email[Corresponding author. Electronic address:\\]{denis\_sheka@univ.kiev.ua}
 \affiliation{National Taras Shevchenko University of Kiev, 03127 Kiev, Ukraine}

\author{Yuri Gaididei}
 \affiliation{Institute for Theoretical Physics, 03143 Kiev, Ukraine}

\author{Franz G.~Mertens}
 \affiliation{Physics Institute, University of Bayreuth, 95440 Bayreuth, Germany}

\date{21.05.07}

%
%

\begin{abstract}
The switching process of the vortex core in a Permalloy nanodisk affected by a
rotating magnetic field is studied theoretically. A detailed description of
magnetization dynamics is obtained by micromagnetic simulations.
\end{abstract}




\maketitle

Artificial mesoscopic magnetic structures provide now a wide testing area for
concepts of nanomagnetism and numerous prospective applications
\cite{Hubert98,Skomski03}. A remarkable example is a vortex state nanodot.
Having nontrivial topological structure on a scale of a nanomagnet, magnetic
vortex is a promising candidate for a high density magnetic storage and high
speed magnetic random access memory \cite{Cowburn02}. The basis of the vortex
statics and dynamics in Heisenberg magnets was studied in 1980s, for a review
see Ref.~\cite{Mertens00}. Typically, the vortex is considered as a rigid
particle without internal degrees of freedom. Using such an approach a number
of dynamical effects were studied in Heisenberg magnets \cite{Mertens00} and
also in nanomagnets \cite{Guslienko02a}. However, the rigid approach fails
when considering the vortex dynamics under the influence of a strong or fast
external force. In particular, it is known that external pumping excites
internal modes in vortex dynamics in Heisenberg magnets, whose role is
important for understanding the switching phenomena
\cite{Gaididei99,Gaididei00,Kovalev02,Kovalev03,Zagorodny03,Caputo07}, and the
limit cycles in the vortex dynamics \cite{Zagorodny04,Sheka05}.

In the present work we study the effect of influence of the homogeneous
rotating field $\vec{B}(t) = \left(B\cos\omega t, B\sin\omega t, 0\right)$ on
nonlinear internal dynamics of the vortex state magnetic nanodot in the
framework of Landau-Lifshitz-Gilbert equations with account of the exchange
and dipolar interaction. Using micromagnetic simulations for material
parameters of Permalloy (Py), \footnote[1]{In all simulations we used material
parameters adopted for the Py particle: the exchange constant
$A=1.3\times10^{-6}$ erg/cm, the saturation magnetization
$M_S=8.6\times10^{2}$ G, the damping coefficient $\alpha = 0.006$ and the
anisotropy was neglected. This corresponds to the exchange length $\ell =
\sqrt{A/4\pi M_S^2} \approx 5.3$nm, which determines the typical radius of the
vortex core. The mesh cells have sizes $3\times3\times20$ nm.} we found that
the irreversible flipping process of the vortex core polarity $p$ is possible
in a specific range of the field parameters $(B,\omega)$, when the direction
of $\vec{\omega}$ is opposite to the vortex polarity, i.e. when $\omega p<0$.

The simplest way to flip the vortex can be realized by applying a
perpendicular DC field $B_\perp$. Due to this field the heavy vortex, which is
polarized against the field, becomes unstable \cite{Ivanov95b,Ivanov02} and at
some critical value ($B_s\sim 1.5 \pi M_S$ \cite{Kravchuk07a}, $M_S$ is the
saturation magnetization \cite{endnote1}) it flips to the light vortex state
in which the core magnetization is parallel to the field. Note that this
switching field should be strong enough, $B_s \sim 2.5$ kOe
\cite{Kikuchi01,Okuno02,Thiaville03,Kravchuk07a}. Another possibility to
switch the vortex is to use a weak but fast rotating field $\vec{B}(t)$
\cite{Gaididei00,Kovalev02,Kovalev03,Zagorodny03}. The physical idea is very
simple. Let us consider our system a moving frame, which rotates with a
frequency $\omega$ together with the field. In this frame instead of the
rotating field one has a static in-plane field $B$. According to the Larmor
theorem there appears also an effective perpendicular field $B_\perp = \hslash
\omega/(g\mu_B)$. Thus the switching problem under the action of the AC field
reduces to the problem with two DC fields. This idea perfectly works for the
Heisenberg magnets and the dynamical picture of the switching is almost the
same for both DC and AC field \cite{Zagorodny03}. However, as we show in this
letter, this simple picture fails for magnetic nanodots, where the dipolar
interaction is decisive for the vortex structure \cite{Hubert98,Skomski03}.

Probably the simplest way which gives physical insight how the dipolar
interaction secures the stability of curling ground state is to use a local
approach \cite{Caputo07a}. In this approach the dipolar interaction can be
reduced approximately to an on-site anisotropy energy which in the case of a
cylindrical nanodot has the form
\begin{equation} \label{eq:dipolar}
\begin{split}
\frac{E_{\text{dipolar}}}{\pi R^2 h M_S^2} \approx &\!\!\int\!\!
\mathrm{d}^2x\! \Bigl[\mathscr{A}(r) + \mathscr{B}(r)\cos2(\phi-\chi) \Bigr]
\sin^2\theta.
\end{split}
\end{equation}
Here ($r,\chi$) are the polar coordinates in the XY plane and we use the
angular parametrization for the magnetization $M_x+\imath M_y =
M_S\sin\theta\exp(\imath\phi)$ and $M_z=M_S\cos\theta$. For thin nanodots the
easy-plane anisotropy parameter $\mathscr{A}(r)$ is negative being practically
a constant. The in-plane anisotropy parameter $\mathscr{B}(r)$ is
coordinate-dependent: it vanishes in the disk center and it is sharply
localized near the disk edge \cite{Caputo07a}. One can see that for the
easy-plane uniform state nanodot with $\phi=\text{const}$, $\theta=\pi/2$, the
second (in-plane) energy term in \eqref{eq:dipolar} vanishes after integration
and makes no important contribution. However, for the vortex state
\begin{equation} \label{eq:vortex}
\cos\theta = p f\left(\frac{r}{\ell}\right), \qquad \phi = q\chi +\aleph
\frac{\pi}{2}
\end{equation}
with $q=1$ being the vorticity and $\aleph = \pm1$ being the chirality,
$\cos2(\phi-\chi)=-1$ and the dipolar interaction \eqref{eq:dipolar} is of
crucial importance. In \eqref{eq:vortex} an exponentially localized function
$f(r/\ell)$ describes the vortex core structure. For thin disks one can
neglect the z-dependence of the magnetization \cite{Kravchuk07a}. The
topological properties of the magnetization distribution are characterized by
the Pontryagin index
\begin{equation} \label{eq:Ponryagin-index}
Q = \frac{1}{4\pi}\int \mathrm{d}^2x\mathcal{Q}, \quad \mathcal{Q} =
\epsilon_{ij}\sin\theta\partial_i\theta \partial_j\phi.
\end{equation}
For the vortex configuration \eqref{eq:vortex} the Pontryagin index is
$Q=pq/2$.

\begin{figure}
\includegraphics[width=85mm]{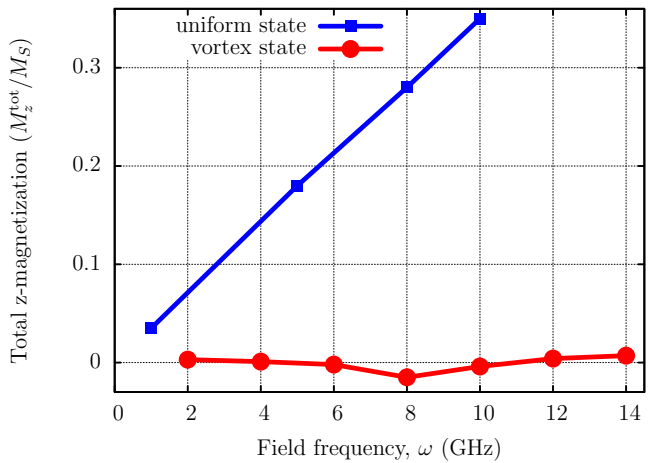}
\caption{(Color online) The ground state magnetization of the nanodot as a
function of the AC field frequency: blue squares correspond to the uniform
state nanodot ($80$ nm diameter, $2$ nm thickness), red circles to the vortex
state nanodot ($132$ nm diameter, $20$ nm thickness).
For both cases the field strength $B=0.05$ T.}%
\label{fig:M_omega}%
\end{figure}

To study the vortex core dynamics in a nanoparticle we performed a detailed
simulations, using the OOMMF micromagnetic simulator for
Landau-Lifshitz-Gilbert equations \cite{OOMMF}. We simulated the vortex state
Py nanodisk \cite{endnote1} with the diameter $2L = 132$ nm and the thickness
$h=20$ nm.

\begin{figure}[ht]
\centerline{$t=1.1$ ns}
\leftline{(a) \hspace{37mm} (b)}
\vspace{-5mm}
\includegraphics[width=44mm]{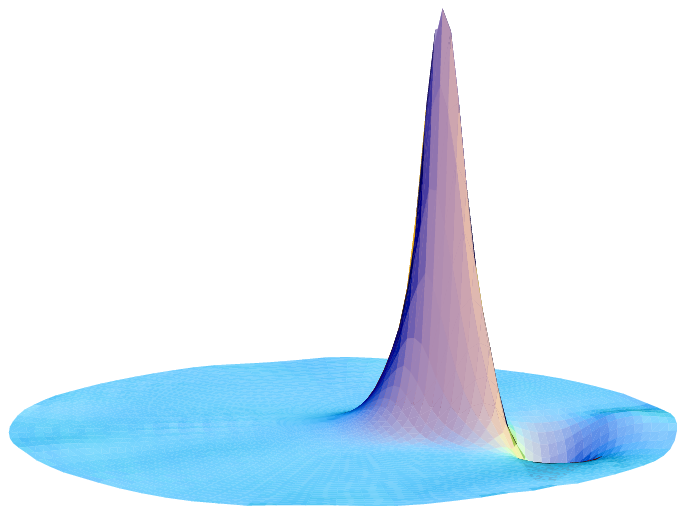}
\includegraphics[width=34mm]{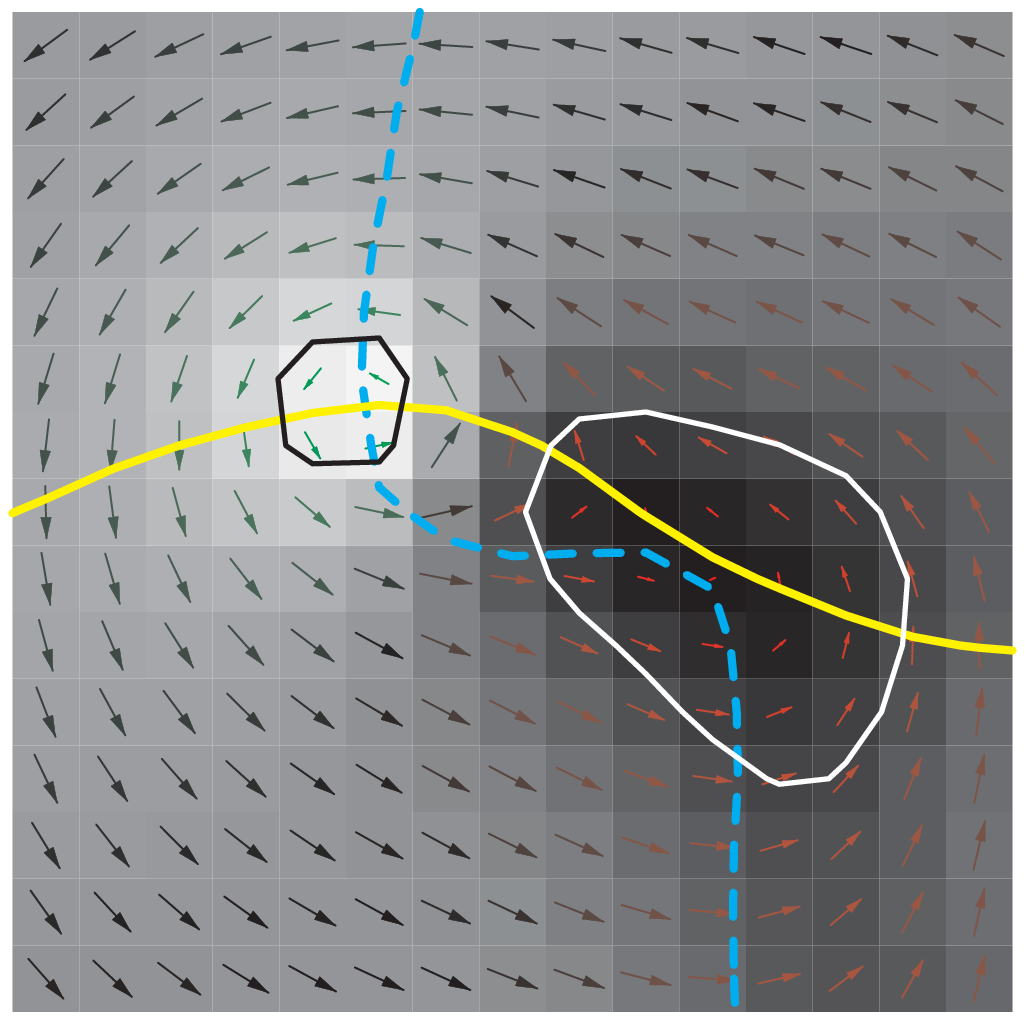}\\[-0.5mm]
\centerline{$t=1.18$ ns}
\leftline{(c) \hspace{37mm} (d)}
\vspace{-5mm}
\includegraphics[width=44mm]{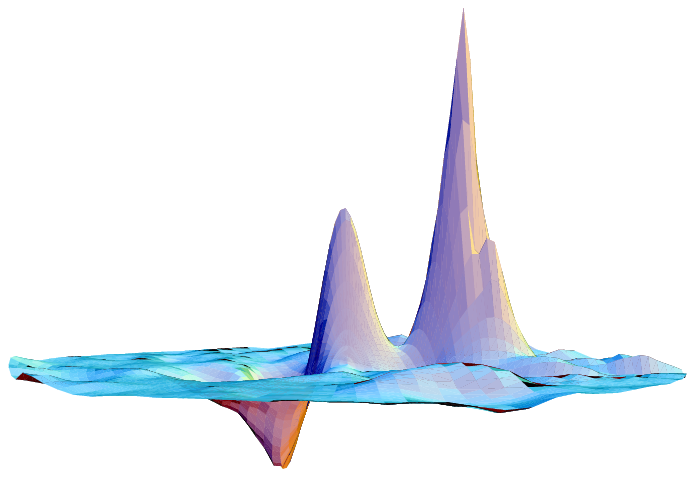}
\includegraphics[width=34mm]{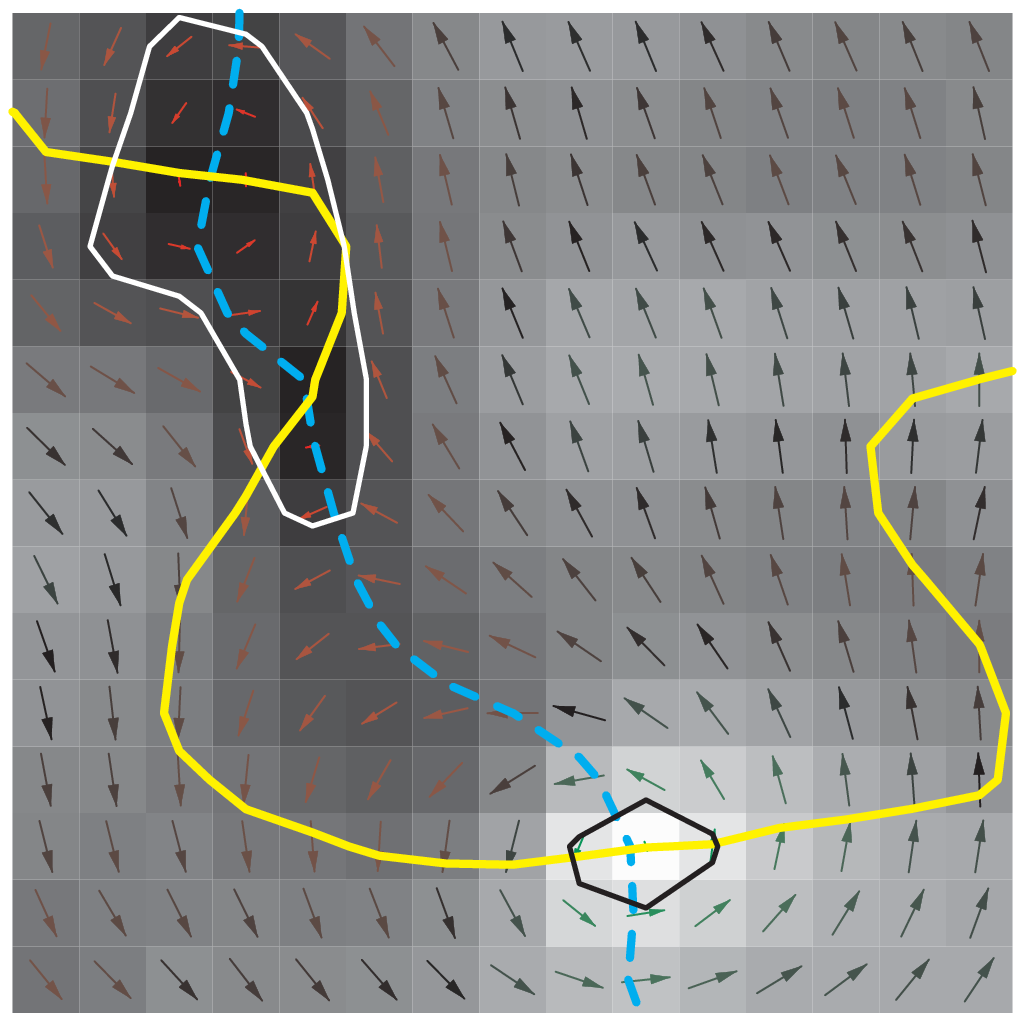}\\[-0.5mm]
\centerline{$t=1.33$ ns}
\leftline{(e) \hspace{37mm} (f)}
\vspace{-5mm}
\includegraphics[width=44mm]{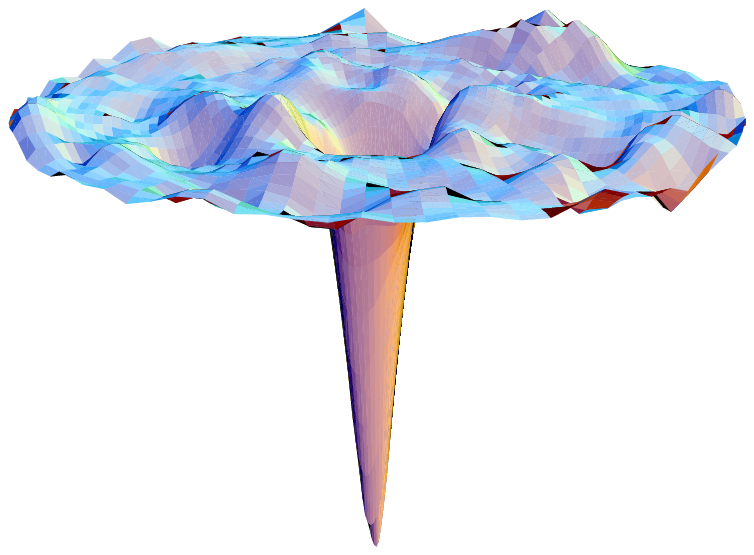}
\includegraphics[width=34mm]{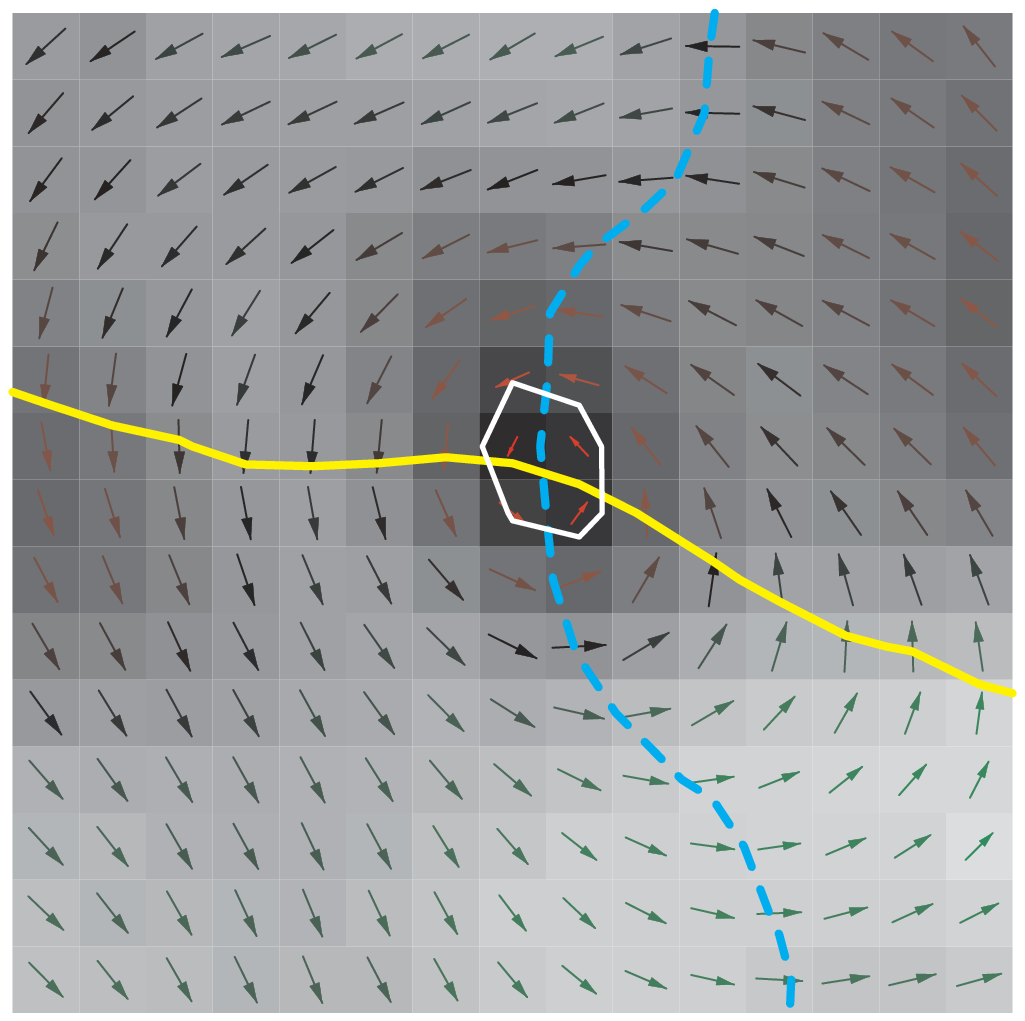}
\caption{(Color online) The temporal picture of the vortex switching process
for the $132$ nm diameter and $20$ nm thickness Py disk: the left panel
[Figs. (a), (c), (e)] corresponds to the 3D distribution of the topological
density $\mathcal{Q}$, see Eq.~\eqref{eq:Ponryagin-index}, and the right panel
[Figs. (b), (d), (f)] represents the in-plane magnetization distribution.
The dashed blue and the solid yellow
curves represent $M_y=0$ and $M_x=0$ isosurfaces, respectively; the black and
the white curves correspond to $M_z/M_S=0.75$ and $M_z/M_S=-0.75$ isosurfaces,
respectively. The field parameters: $\omega=4$ GHz and $B=0.06$ T.}
\label{fig:switching}%
\end{figure}

\begin{figure*}
\includegraphics[width=44mm]{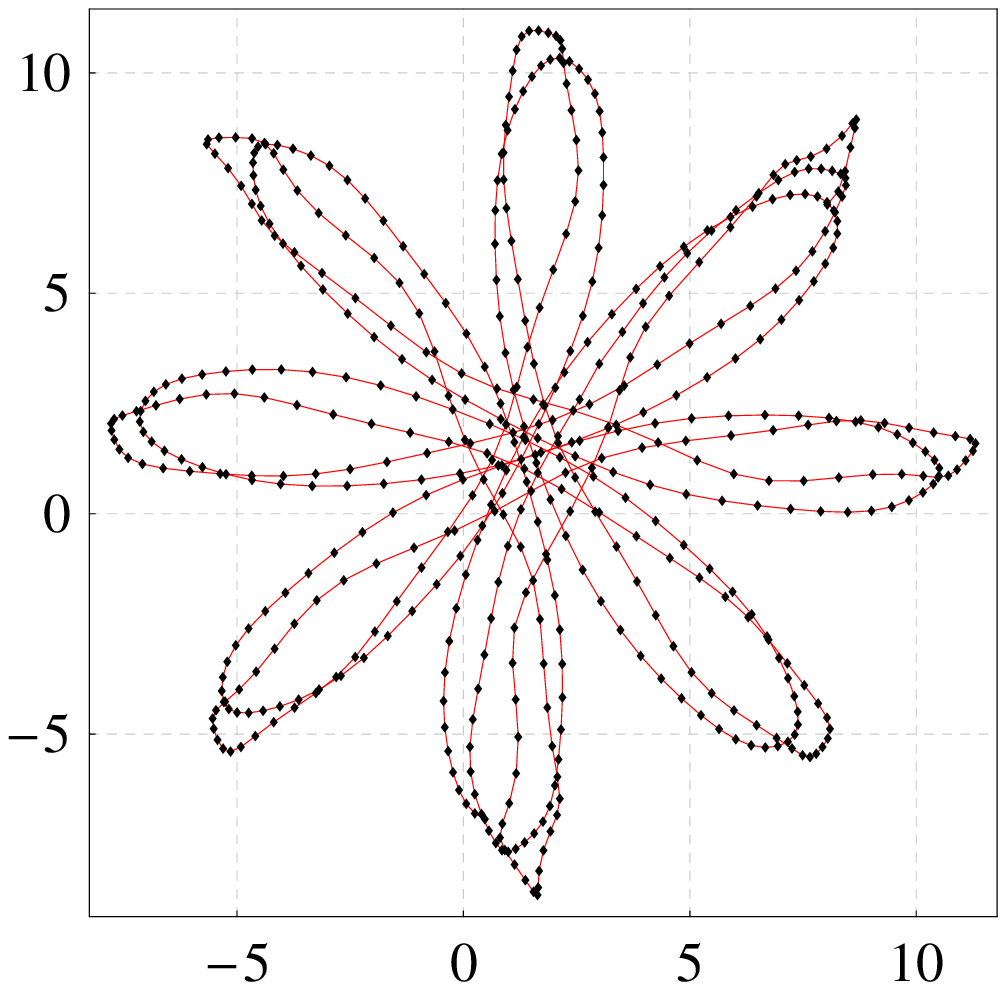}
\includegraphics[width=44mm]{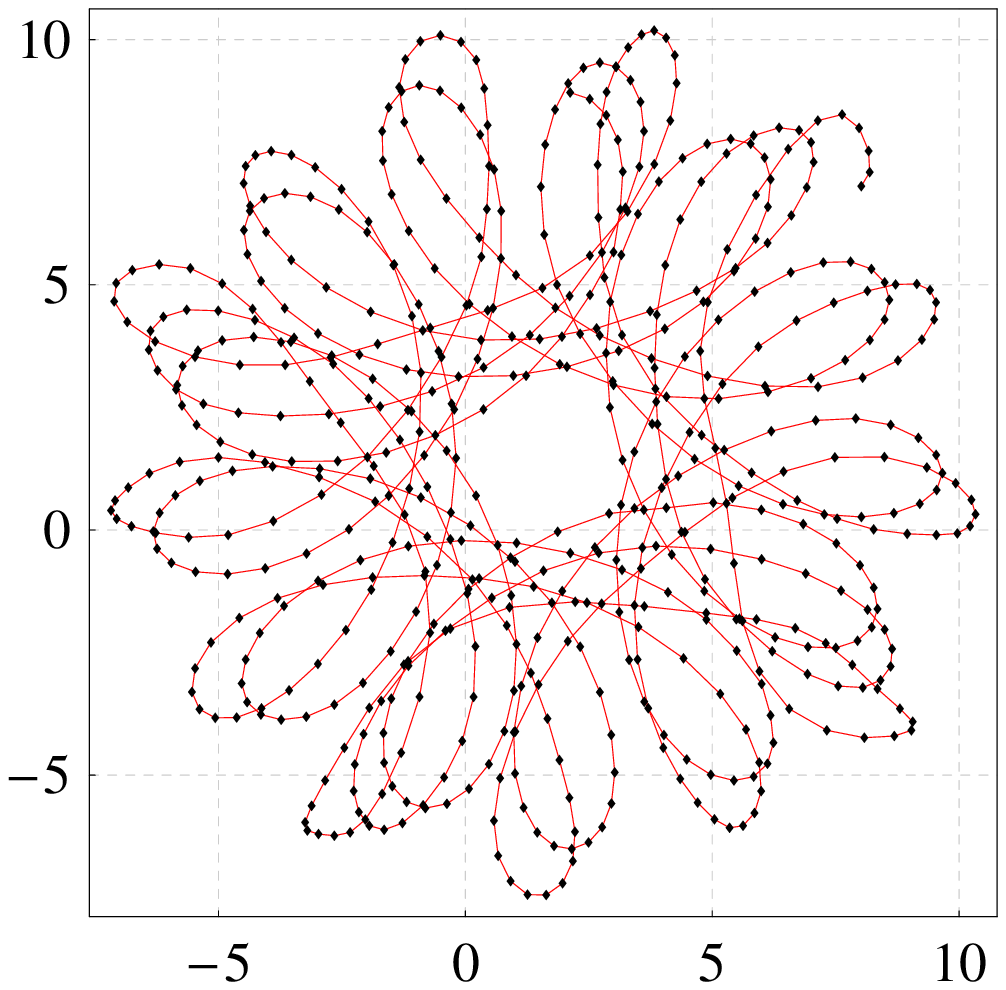}
\includegraphics[width=44mm]{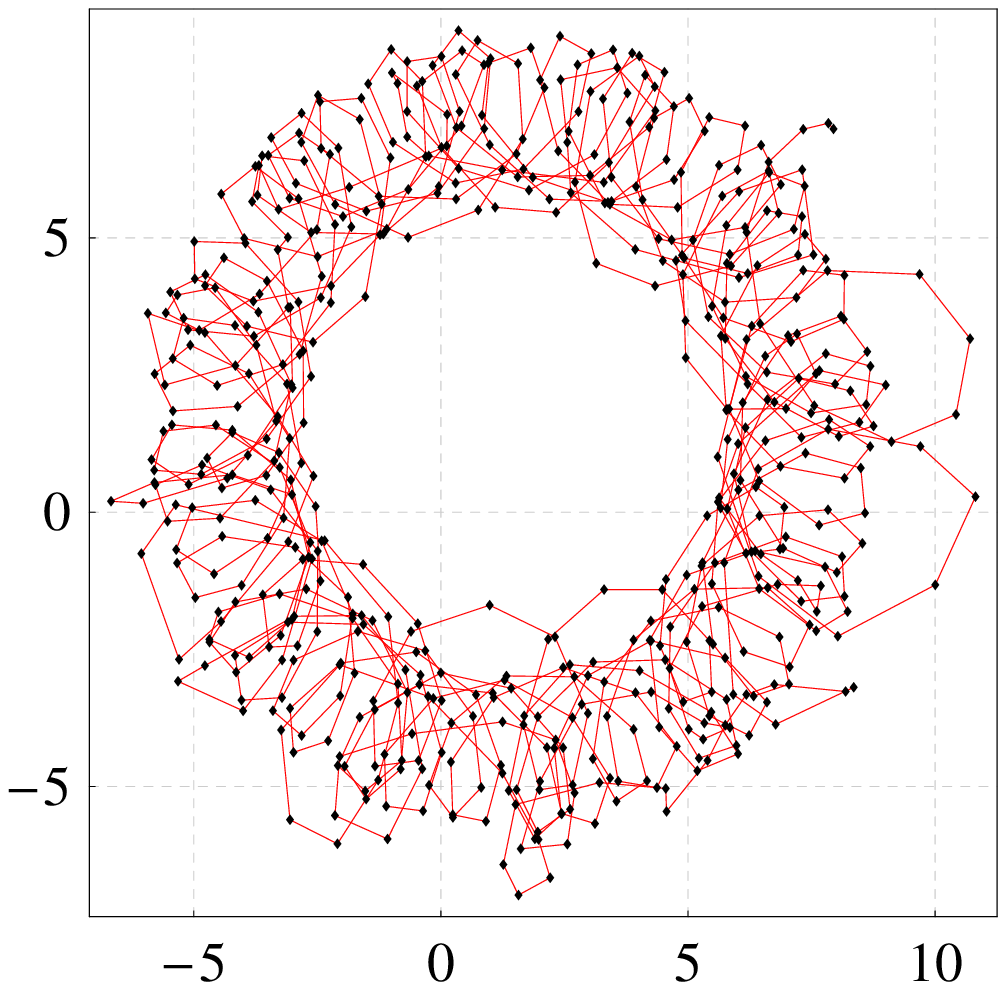}
\includegraphics[width=44mm]{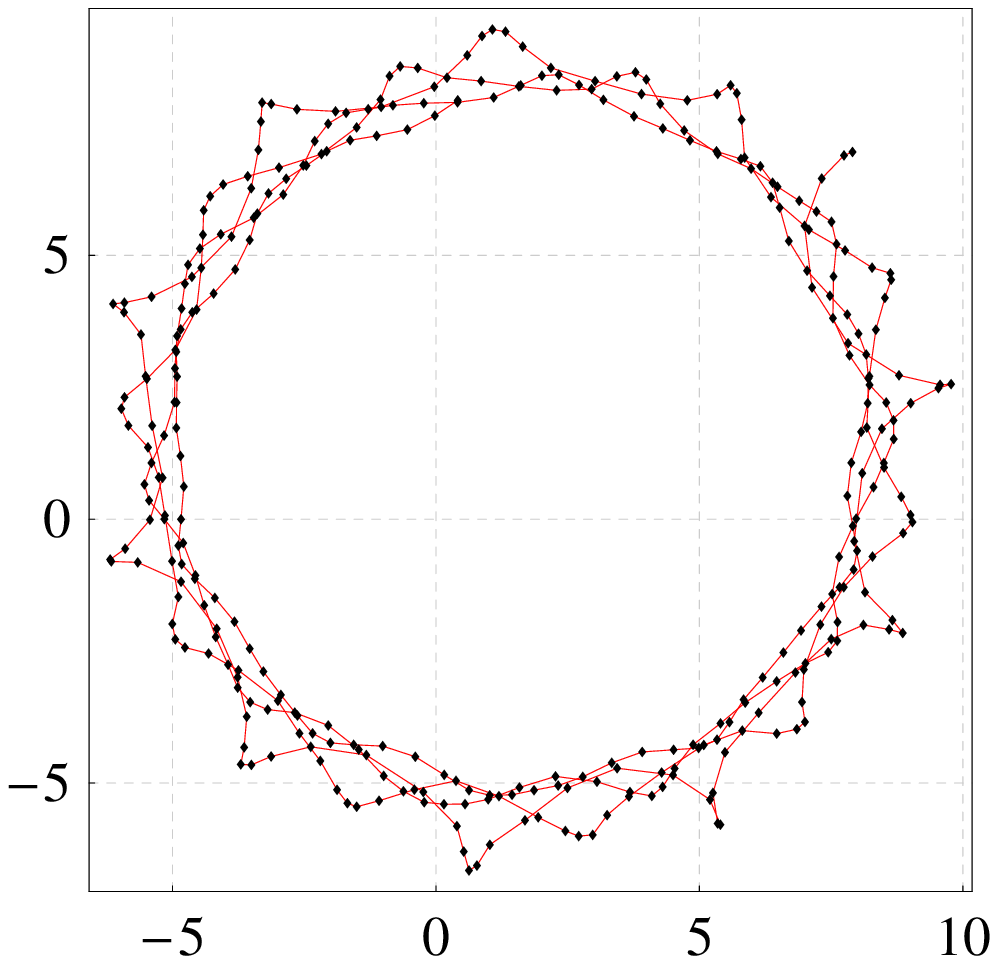}\\
\centerline{%
\                  \hfill %
(a) $\omega=2$ GHz \hfill %
(b) $\omega=4$ GHz \hfill %
(c) $\omega=14$ GHz \hfill %
(d) $\omega=20$ GHz \hspace{10mm}%
}%
\caption{(Color online) Vortex trajectories (measured in nanometers) for
different frequencies. The field amplitude always is the same (0.02T).}
\label{fig:trajectory}%
\end{figure*}

First we decided to validate the analogy between the rotating AC field and
effective DC field for the nanodot. Since the visual effect of the DC field is
the change of the ground state magnetization, we calculated the dependence of
$M_z^{\text{tot}}$ on the frequency of the AC field $\omega$, see
Fig.~\ref{fig:M_omega}. For the uniform state nanodot, the dependence is
linear, $M_z^{\text{tot}} = \hslash\omega/(4\pi g\mu_B)$, hence the analogy
with the effective static field $B_\perp$ works correctly for this case.
However it fails for the vortex state nanodot, where $M_z^{\text{tot}}\approx
0$, see Fig.~\ref{fig:M_omega}. The reason for this is the following. While in
infinite systems the transformation to the rotating frame of reference in the
space of spin variables ($\phi\to\tilde{\phi}=\phi+\omega t$) makes the
Hamiltonian of the system time independent, this is not the case in
finite-size systems. As is seen from \eqref{eq:dipolar} (and takes place also
for the total dipole-dipole interaction) to produce a time independent
Hamiltonian in addition to the spin transformation one should transfer to the
rotating frame of reference in real space: $\chi\to\tilde{\chi}=\chi+\omega
t$. Thus in the rotating frame one has a static magnetic field $\vec{B}$ along
one axis in a disk plane. Besides that, taking into account that the
Lagrangian of the system contains a gyroscopical term of the form $\int
\mathrm{d}^2x(p-\cos\theta)\partial_t{\phi}$, \cite{Sheka06e} the Larmor and
gyroscopical terms provide an additional energy contribution
\begin{equation} \label{eq:E-eff}
\Delta E = \frac{M_S h\omega}{\gamma}\int \mathrm{d}^2\tilde{x}
(p-\cos\theta)\left(1-\frac{\partial\tilde{\phi}}{
\partial{\tilde{\chi}}}\right).
\end{equation}
Without a vortex, the additional energy term \eqref{eq:E-eff} is equivalent to
the action of an effective perpendicular magnetic field, $\Delta E = -B_\perp
M_z^{\text{tot}}$, which results in a not vanishing $M_z^{\text{tot}}$, in
accordance with above results, see Fig.~\ref{fig:M_omega}. If the vortex is
present, the situation drastically changes. In particular, $\Delta E=0$ for
the vortex which is situated exactly in a disk center. This explains why we do
no have an essential field influence on a centered vortex.

If we apply the field, which rotates clockwise ($\omega<0$), the vortex with
$p=+1$ starts to move counterclockwise. This vortex motion has a gyroscopic
nature: the vortex trajectory $\vec{R}(t)$ results mainly from the balance
between the field interaction and the internal gyroscopic force $\vec{G}
\times \mathrm{d}\vec{R}/\mathrm{d}t$, where $\vec{G} = -4\pi M_S h
\gamma^{-1} Q\vec{e}_z$ is the gyrovector.

During its motion, the vortex excites a number of spin waves \cite{Choi07},
the vortex trajectory becomes complicated and sometimes it is tricky to
identify the exact vortex position. We used the spatial distribution
\eqref{eq:Ponryagin-index} of the topological density $\mathcal{Q}$,  to
determine the vortex center position $\vec{R}(t)$. This distribution of
$\mathcal{Q}$ is shown on the left panel of Fig.~\ref{fig:switching}. Typical
vortex trajectories are shown in Fig.~\ref{fig:trajectory}. Mainly two magnon
modes are excited in this system: symmetrical and azimuthal ones
\cite{Gaididei00,Zagorodny03}. The azimuthal mode is caused by the influence
of the in-plane field and the symmetrical one by the effective out-of plane
field $B_\perp$. Due to the continuous pumping the system goes to the
nonlinear regime: the amplitude of the azimuthal mode increases and there
appears an out-of-plane dip nearby the vortex, similar to the vortex structure
distortion by the in-plane field pulse \cite{Hertel07,Waeyenberge06,Xiao06} or
an AC oscillating field \cite{Lee07}; this dip is shown on the right panel of
Fig.~\ref{fig:switching} by the dark gray color and it is surrounded by the
white isosurface. To highlight the vortex core position on the XY plane (right
panel of Fig.~\ref{fig:switching}), we plot the isosurfaces $M_x=0$ and
$M_y=0$: each intersection of these isosurfaces determines the position of the
vortex ($q=+1$) or the antivortex ($q=-1$) \cite{Hertel07}. All isosurfaces
are closed inside the disk, or go the the disk edge, hence the number of
crossing points can be changed during the system dynamics only by two,
reflecting the processes of the vortex-antivortex pair creation and
annihilation.

\begin{figure}[b]
\includegraphics[width=85mm]{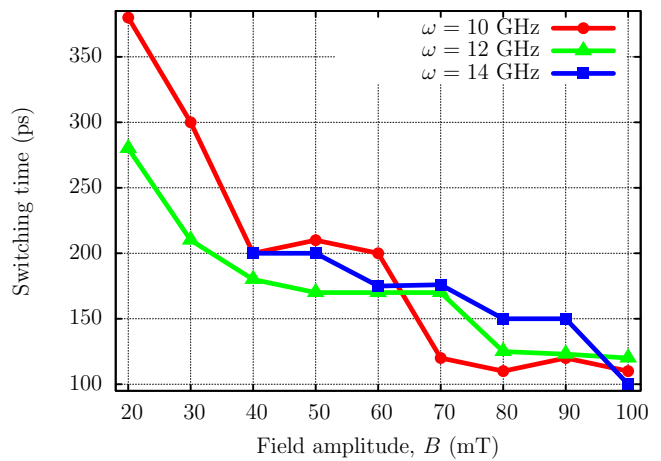}
\caption{(Color online) The switching time for different field parameters.
The size parameters of the dots are the same as in Fig.~\ref{fig:switching}}%
\label{fig:sw_time}%
\end{figure}

When the amplitude of the out-of-plane dip reaches the minimum ($M_z=-M_S$) a
pair of a \emph{new} vortex ($p=-1$, $q=+1$) and antivortex ($p=-1$, $q=-1$)
is created, see Figs.~\ref{fig:switching} (c) and (d). Since the total
topological charge of the pair $Q=0$, this process does not cost a lot of
energy \cite{Tretiakov07}. On Fig.~\ref{fig:switching}(c) one can see three
peaks in the $\mathcal{Q}$-distribution. Two positive peaks correspond to the
initial vortex and the new antivortex. The negative peak corresponds to the
new vortex. After that the scenario of the vortex-vortex-antivortex dynamics
is similar to the one discussed by \citet{Waeyenberge06}. The original vortex
and the new antivortex interact with each other. In the exchange limit
(neglecting the dipolar interaction) the vortex-antivortex pair is
topologically equivalent to the Belavin-Polyakov soliton \cite{Belavin75} with
the topological charge $Q=1$. This soliton is scale invariant in the continuum
system. However, in the discrete system the radius of this soliton, i.e. the
distance between the vortex and antivortex, rapidly decreases almost without
energy lost, since this soliton is not stable. When the soliton radius is
about the lattice constant, it can vanishes; this vanishing is accompanied by
strong spin-wave radiation, because the topological properties of the system
change \cite{Hertel06,Tretiakov07}. After annihilation of the
vortex-antivortex pair only the new vortex ($p=-1$) survives, see
Figs.~\ref{fig:switching} (e) and (f). After the flipping the vortex changes
the direction of motion, because it has now the opposite polarity. The
interaction with the rotating field is not so strong now, and the vortex
rapidly goes to the disk center on a spiral trajectory. Thus, using the
rotating field we obtain an \emph{irreversible} switching of the vortex
polarity. The Supplementary Video illustrates the whole evolution of the
magnetization distribution during the switching process
\footnote[2]{Supplemental video,
\url{http://www.phy.uni-bayreuth.de/\~btp315/publ/Kravchuk.arXiv.0705.2046.html}}.
Typically the switching time is in the hundreds ps range; it can be decreased
by more intensive fields, see Fig.~\ref{fig:sw_time} \footnote[3]{In our
simulations the vortex was initially shifted from the disk center on a
distance about its core radius. The switching time slightly increases when we
place the vortex exactly into the disk center.}.

\begin{figure}
\includegraphics[width=85mm]{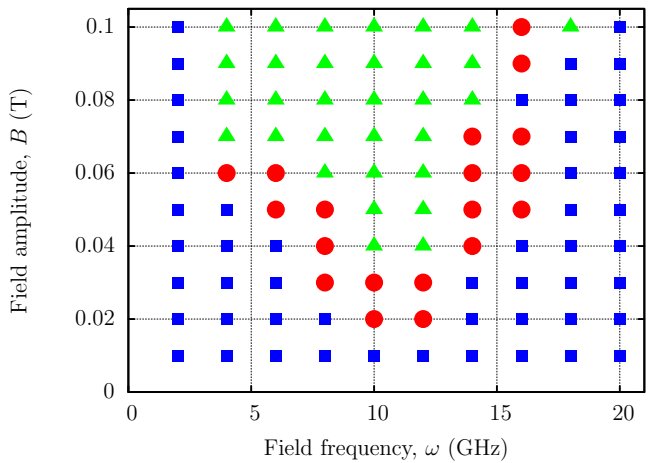}
\caption{(Color online) Switching diagram for the ac field parameters
(amplitude and frequency): The symbols {\huge \textcolor{red}{$\bullet$}}
correspond to parameters, when the switching occurs,
\textcolor{blue}{$\blacksquare$} describe the vortex dynamics without
switching, and {\Large \textcolor{green}{$\blacktriangle$}} represent
multiple switching processes.}%
\label{fig:switch_diagr}%
\end{figure}

The switching occurs in a well defined range of the field parameters
($\omega,B$) with the minimum frequency about $10$ GHz and the field strength
about $20$ mT, see Fig.~\ref{fig:switch_diagr}. When the field intensity is
not strong enough, the vortex moves without switching. In the opposite case of
strong fields, multiple switching occurs, similar to \cite{Hertel07,Lee07}.
When the frequency is too large or too small the switching also does not
happen. Thus, the controlled switching process can be realized experimentally
only for a relatively narrow range of parameters (red disks in the switching
diagram, Fig.~\ref{fig:switch_diagr}). The switching diagram differs from the
one in Heisenberg magnets, where the vortex flipping is characterized mainly
by a threshold curve $B_{\text{cr}}(\omega)$ \cite{Gaididei00,Zagorodny03}.

Note that in the range of the high frequency pumping, when the switching does
not occur (the right sight of the diagram \ref{fig:switch_diagr}), the vortex
always reaches some limit cycle, which is almost a circular trajectory, see
Fig.~\ref{fig:trajectory}(d). An important feature of this limit cycle is that
the frequency of the vortex orbital motion $\Omega\sim 1$ GHz is much less
than the pumping frequency $\omega$ and almost does not depend on the field
frequency and its amplitude. We expect that this low frequency response
results from the coupling between the vortex motion and internal modes
dynamics in analogue with the Heisenberg magnets \cite{Zagorodny04,Sheka05}.
The detailed analysis of this vortex dynamics is under construction.

In conclusion, we have described the vortex core dynamics in the nanodisk
under periodic driving. The irreversible switching of the vortex polarity can
be excited by a high-frequency ($10$ GHz) and intensity ($20$ mT) AC field.
The mechanism of the switching involving the creation and annihilation of a
vortex-antivortex pair is quite general. It appears for different kinds of
vortex excitation and is induced in particular by a field pulse
\cite{Waeyenberge06,Xiao06,Hertel06a}, by an AC oscillating field
\cite{Lee07}, and even by an electrical current \cite{Yamada07,Liu07}. We also
reported about the limit cycle in the vortex dynamics for the higher frequency
pumping. Both effects are caused by the strong coupling with excited internal
magnon modes.

\acknowledgments

The authors thank S. Komineas, H. Stoll and M. F\"{a}hnle for helpful
discussions. V.K., D.S., Yu.G. thank the University of Bayreuth, where this
work was performed, for kind hospitality and acknowledge the support from DLR
grant No.~UKR~05/055. V.K. acknowledges the support from the BAYHOST project.
D.S. acknowledges the support from the Alexander von Humboldt--Foundation.


\begin{thebibliography}{33}
\expandafter\ifx\csname natexlab\endcsname\relax\def\natexlab#1{#1}\fi
\expandafter\ifx\csname bibnamefont\endcsname\relax
  \def\bibnamefont#1{#1}\fi
\expandafter\ifx\csname bibfnamefont\endcsname\relax
  \def\bibfnamefont#1{#1}\fi
\expandafter\ifx\csname citenamefont\endcsname\relax
  \def\citenamefont#1{#1}\fi
\expandafter\ifx\csname url\endcsname\relax
  \def\url#1{\texttt{#1}}\fi
\expandafter\ifx\csname urlprefix\endcsname\relax\def\urlprefix{URL }\fi
\providecommand{\bibinfo}[2]{#2}
\providecommand{\eprint}[2][]{\url{#2}}

\bibitem[{\citenamefont{Hubert and Sch{\" a}fer}(1998)}]{Hubert98}
\bibinfo{author}{\bibfnamefont{A.}~\bibnamefont{Hubert}} \bibnamefont{and}
  \bibinfo{author}{\bibfnamefont{R.}~\bibnamefont{Sch{\" a}fer}},
  \emph{\bibinfo{title}{Magnetic domains}}
  (\bibinfo{publisher}{Springer--Verlag}, \bibinfo{address}{Berlin},
  \bibinfo{year}{1998}).

\bibitem[{\citenamefont{Skomski}(2003)}]{Skomski03}
\bibinfo{author}{\bibfnamefont{R.}~\bibnamefont{Skomski}},
  \bibinfo{journal}{J.~Phys.} \textbf{\bibinfo{volume}{C 15}},
  \bibinfo{pages}{R841} (\bibinfo{year}{2003}),
  \urlprefix\url{http://www.iop.org/EJ/abstract/0953-8984/15/20/202/}.

\bibitem[{\citenamefont{Cowburn}(2002)}]{Cowburn02}
\bibinfo{author}{\bibfnamefont{R.~P.} \bibnamefont{Cowburn}},
  \bibinfo{journal}{J.~Magn. Magn. Mater.} \textbf{\bibinfo{volume}{242-245}},
  \bibinfo{pages}{505} (\bibinfo{year}{2002}),
  \urlprefix\url{http://www.sciencedirect.com/science/article/B6TJJ-44NM486-19%
/2/d76afadecc91907bb34369038cc8368e}.

\bibitem[{\citenamefont{Mertens and Bishop}(2000)}]{Mertens00}
\bibinfo{author}{\bibfnamefont{F.~G.} \bibnamefont{Mertens}} \bibnamefont{and}
  \bibinfo{author}{\bibfnamefont{A.~R.} \bibnamefont{Bishop}}, in
  \emph{\bibinfo{booktitle}{Nonlinear Science at the Dawn of the 21th
  Century}}, edited by \bibinfo{editor}{\bibfnamefont{P.~L.}
  \bibnamefont{Christiansen}}, \bibinfo{editor}{\bibfnamefont{M.~P.}
  \bibnamefont{Soerensen}}, \bibnamefont{and}
  \bibinfo{editor}{\bibfnamefont{A.~C.} \bibnamefont{Scott}}
  (\bibinfo{publisher}{Springer--Verlag}, \bibinfo{address}{Berlin},
  \bibinfo{year}{2000}).

\bibitem[{\citenamefont{Guslienko et~al.}(2002)\citenamefont{Guslienko, Ivanov,
  Novosad, Otani, Shima, and Fukamichi}}]{Guslienko02a}
\bibinfo{author}{\bibfnamefont{K.~Y.} \bibnamefont{Guslienko}},
  \bibinfo{author}{\bibfnamefont{B.~A.} \bibnamefont{Ivanov}},
  \bibinfo{author}{\bibfnamefont{V.}~\bibnamefont{Novosad}},
  \bibinfo{author}{\bibfnamefont{Y.}~\bibnamefont{Otani}},
  \bibinfo{author}{\bibfnamefont{H.}~\bibnamefont{Shima}}, \bibnamefont{and}
  \bibinfo{author}{\bibfnamefont{K.}~\bibnamefont{Fukamichi}},
  \bibinfo{journal}{J.~Appl. Phys.} \textbf{\bibinfo{volume}{91}},
  \bibinfo{pages}{8037} (\bibinfo{year}{2002}),
  \urlprefix\url{http://link.aip.org/link/?JAP/91/8037/1}.

\bibitem[{\citenamefont{Gaididei et~al.}(1999)\citenamefont{Gaididei,
  Kamppeter, Mertens, and Bishop}}]{Gaididei99}
\bibinfo{author}{\bibfnamefont{Y.}~\bibnamefont{Gaididei}},
  \bibinfo{author}{\bibfnamefont{T.}~\bibnamefont{Kamppeter}},
  \bibinfo{author}{\bibfnamefont{F.~G.} \bibnamefont{Mertens}},
  \bibnamefont{and} \bibinfo{author}{\bibfnamefont{A.}~\bibnamefont{Bishop}},
  \bibinfo{journal}{Phys. Rev. B} \textbf{\bibinfo{volume}{59}},
  \bibinfo{pages}{7010} (\bibinfo{year}{1999}),
  \urlprefix\url{http://link.aps.org/abstract/PRB/v59/p7010}.

\bibitem[{\citenamefont{Gaididei et~al.}(2000)\citenamefont{Gaididei,
  Kamppeter, Mertens, and Bishop}}]{Gaididei00}
\bibinfo{author}{\bibfnamefont{Y.}~\bibnamefont{Gaididei}},
  \bibinfo{author}{\bibfnamefont{T.}~\bibnamefont{Kamppeter}},
  \bibinfo{author}{\bibfnamefont{F.~G.} \bibnamefont{Mertens}},
  \bibnamefont{and} \bibinfo{author}{\bibfnamefont{A.~R.}
  \bibnamefont{Bishop}}, \bibinfo{journal}{Phys. Rev. B}
  \textbf{\bibinfo{volume}{61}}, \bibinfo{pages}{9449} (\bibinfo{year}{2000}),
  \urlprefix\url{http://link.aps.org/abstract/PRB/v61/p9449}.

\bibitem[{\citenamefont{Kovalev and Prilepsky}(2002)}]{Kovalev02}
\bibinfo{author}{\bibfnamefont{A.~S.} \bibnamefont{Kovalev}} \bibnamefont{and}
  \bibinfo{author}{\bibfnamefont{J.~E.} \bibnamefont{Prilepsky}},
  \bibinfo{journal}{Low Temp.~Phys.} \textbf{\bibinfo{volume}{28}},
  \bibinfo{pages}{921} (\bibinfo{year}{2002}).

\bibitem[{\citenamefont{Kovalev and Prilepsky}(2003)}]{Kovalev03}
\bibinfo{author}{\bibfnamefont{A.~S.} \bibnamefont{Kovalev}} \bibnamefont{and}
  \bibinfo{author}{\bibfnamefont{J.~E.} \bibnamefont{Prilepsky}},
  \bibinfo{journal}{Low Temp.~Phys.} \textbf{\bibinfo{volume}{29}},
  \bibinfo{pages}{55} (\bibinfo{year}{2003}).

\bibitem[{\citenamefont{Zagorodny et~al.}(2003)\citenamefont{Zagorodny,
  Gaididei, Mertens, and Bishop}}]{Zagorodny03}
\bibinfo{author}{\bibfnamefont{J.~P.} \bibnamefont{Zagorodny}},
  \bibinfo{author}{\bibfnamefont{Y.}~\bibnamefont{Gaididei}},
  \bibinfo{author}{\bibfnamefont{F.~G.} \bibnamefont{Mertens}},
  \bibnamefont{and} \bibinfo{author}{\bibfnamefont{A.~R.}
  \bibnamefont{Bishop}}, \bibinfo{journal}{Eur.~Phys.~J.}
  \textbf{\bibinfo{volume}{B 31}}, \bibinfo{pages}{471} (\bibinfo{year}{2003}),
  \urlprefix\url{http://dx.doi.org/10.1140/epjb/e2003-00057-y}.

\bibitem[{\citenamefont{Caputo et~al.}(2007{\natexlab{a}})\citenamefont{Caputo,
  Gaididei, Mertens, and Sheka}}]{Caputo07}
\bibinfo{author}{\bibfnamefont{J.-G.} \bibnamefont{Caputo}},
  \bibinfo{author}{\bibfnamefont{Y.}~\bibnamefont{Gaididei}},
  \bibinfo{author}{\bibfnamefont{F.~G.} \bibnamefont{Mertens}},
  \bibnamefont{and} \bibinfo{author}{\bibfnamefont{D.~D.} \bibnamefont{Sheka}},
  \bibinfo{journal}{Phys. Rev. Lett.} \textbf{\bibinfo{volume}{98}},
  \bibinfo{eid}{056604} (pages~\bibinfo{numpages}{4})
  (\bibinfo{year}{2007}{\natexlab{a}}),
  \urlprefix\url{http://link.aps.org/abstract/PRL/v98/e056604}.

\bibitem[{\citenamefont{Zagorodny et~al.}(2004)\citenamefont{Zagorodny,
  Gaididei, Sheka, Caputo, and Mertens}}]{Zagorodny04}
\bibinfo{author}{\bibfnamefont{J.~P.} \bibnamefont{Zagorodny}},
  \bibinfo{author}{\bibfnamefont{Y.}~\bibnamefont{Gaididei}},
  \bibinfo{author}{\bibfnamefont{D.~D.} \bibnamefont{Sheka}},
  \bibinfo{author}{\bibfnamefont{J.-G.} \bibnamefont{Caputo}},
  \bibnamefont{and} \bibinfo{author}{\bibfnamefont{F.~G.}
  \bibnamefont{Mertens}}, \bibinfo{journal}{Phys. Rev. Lett.}
  \textbf{\bibinfo{volume}{93}}, \bibinfo{eid}{167201}
  (pages~\bibinfo{numpages}{4}) (\bibinfo{year}{2004}),
  \urlprefix\url{http://link.aps.org/abstract/PRL/v93/e167201}.

\bibitem[{\citenamefont{Sheka et~al.}(2005)\citenamefont{Sheka, Zagorodny,
  Caputo, Gaididei, and Mertens}}]{Sheka05}
\bibinfo{author}{\bibfnamefont{D.~D.} \bibnamefont{Sheka}},
  \bibinfo{author}{\bibfnamefont{J.~P.} \bibnamefont{Zagorodny}},
  \bibinfo{author}{\bibfnamefont{J.~G.} \bibnamefont{Caputo}},
  \bibinfo{author}{\bibfnamefont{Y.}~\bibnamefont{Gaididei}}, \bibnamefont{and}
  \bibinfo{author}{\bibfnamefont{F.~G.} \bibnamefont{Mertens}},
  \bibinfo{journal}{Phys. Rev. B} \textbf{\bibinfo{volume}{71}},
  \bibinfo{eid}{134420} (pages~\bibinfo{numpages}{15}) (\bibinfo{year}{2005}),
  \urlprefix\url{http://link.aps.org/abstract/PRB/v71/e134420}.

\bibitem[{\citenamefont{Ivanov and Sheka}(1995)}]{Ivanov95b}
\bibinfo{author}{\bibfnamefont{B.~A.} \bibnamefont{Ivanov}} \bibnamefont{and}
  \bibinfo{author}{\bibfnamefont{D.~D.} \bibnamefont{Sheka}},
  \bibinfo{journal}{Low Temp. Phys.} \textbf{\bibinfo{volume}{21}},
  \bibinfo{pages}{881} (\bibinfo{year}{1995}),
  \urlprefix\url{http://link.aip.org/link/?LTP/21/881/1}.

\bibitem[{\citenamefont{Ivanov and Wysin}(2002)}]{Ivanov02}
\bibinfo{author}{\bibfnamefont{B.~A.} \bibnamefont{Ivanov}} \bibnamefont{and}
  \bibinfo{author}{\bibfnamefont{G.~M.} \bibnamefont{Wysin}},
  \bibinfo{journal}{Phys. Rev. B} \textbf{\bibinfo{volume}{65}},
  \bibinfo{eid}{134434} (pages~\bibinfo{numpages}{17}) (\bibinfo{year}{2002}),
  \urlprefix\url{http://link.aps.org/abstract/PRB/v65/e134434}.

\bibitem[{\citenamefont{Kravchuk and Sheka}(2007)}]{Kravchuk07a}
\bibinfo{author}{\bibfnamefont{V.~P.} \bibnamefont{Kravchuk}} \bibnamefont{and}
  \bibinfo{author}{\bibfnamefont{D.~D.} \bibnamefont{Sheka}},
  \bibinfo{journal}{Physics of the Solid State} \textbf{\bibinfo{volume}{49}},
  \bibinfo{pages}{(accepted)} (\bibinfo{year}{2007}).

\bibitem[{\citenamefont{Kikuchi et~al.}(2001)\citenamefont{Kikuchi, Okamoto,
  Kitakami, Shimada, Kim, Otani, and Fukamichi}}]{Kikuchi01}
\bibinfo{author}{\bibfnamefont{N.}~\bibnamefont{Kikuchi}},
  \bibinfo{author}{\bibfnamefont{S.}~\bibnamefont{Okamoto}},
  \bibinfo{author}{\bibfnamefont{O.}~\bibnamefont{Kitakami}},
  \bibinfo{author}{\bibfnamefont{Y.}~\bibnamefont{Shimada}},
  \bibinfo{author}{\bibfnamefont{S.~G.} \bibnamefont{Kim}},
  \bibinfo{author}{\bibfnamefont{Y.}~\bibnamefont{Otani}}, \bibnamefont{and}
  \bibinfo{author}{\bibfnamefont{K.}~\bibnamefont{Fukamichi}},
  \bibinfo{journal}{J.~Appl. Phys.} \textbf{\bibinfo{volume}{90}},
  \bibinfo{pages}{6548} (\bibinfo{year}{2001}),
  \urlprefix\url{http://link.aip.org/link/?JAP/90/6548/1}.

\bibitem[{\citenamefont{Okuno et~al.}(2002)\citenamefont{Okuno, Shigeto, Ono,
  Mibu, and Shinjo}}]{Okuno02}
\bibinfo{author}{\bibfnamefont{T.}~\bibnamefont{Okuno}},
  \bibinfo{author}{\bibfnamefont{K.}~\bibnamefont{Shigeto}},
  \bibinfo{author}{\bibfnamefont{T.}~\bibnamefont{Ono}},
  \bibinfo{author}{\bibfnamefont{K.}~\bibnamefont{Mibu}}, \bibnamefont{and}
  \bibinfo{author}{\bibfnamefont{T.}~\bibnamefont{Shinjo}},
  \bibinfo{journal}{J.~Magn. Magn. Mater.} \textbf{\bibinfo{volume}{240}},
  \bibinfo{pages}{1} (\bibinfo{year}{2002}),
  \urlprefix\url{http://www.sciencedirect.com/science/article/B6TJJ-447DCMV-3/%
2/84e9fdfbf9e0cab3d3182fc4db4b4032}.

\bibitem[{\citenamefont{Thiaville et~al.}(2003)\citenamefont{Thiaville, Garcia,
  Dittrich, Miltat, and Schrefl}}]{Thiaville03}
\bibinfo{author}{\bibfnamefont{A.}~\bibnamefont{Thiaville}},
  \bibinfo{author}{\bibfnamefont{J.~M.} \bibnamefont{Garcia}},
  \bibinfo{author}{\bibfnamefont{R.}~\bibnamefont{Dittrich}},
  \bibinfo{author}{\bibfnamefont{J.}~\bibnamefont{Miltat}}, \bibnamefont{and}
  \bibinfo{author}{\bibfnamefont{T.}~\bibnamefont{Schrefl}},
  \bibinfo{journal}{Phys. Rev. B} \textbf{\bibinfo{volume}{67}},
  \bibinfo{eid}{094410} (pages~\bibinfo{numpages}{12}) (\bibinfo{year}{2003}),
  \urlprefix\url{http://link.aps.org/abstract/PRB/v67/e094410}.

\bibitem[{\citenamefont{Caputo et~al.}(2007{\natexlab{b}})\citenamefont{Caputo,
  Gaididei, Kravchuk, Mertens, and Sheka}}]{Caputo07a}
\bibinfo{author}{\bibfnamefont{J.-G.} \bibnamefont{Caputo}},
  \bibinfo{author}{\bibfnamefont{Y.}~\bibnamefont{Gaididei}},
  \bibinfo{author}{\bibfnamefont{V.~P.} \bibnamefont{Kravchuk}},
  \bibinfo{author}{\bibfnamefont{F.~G.} \bibnamefont{Mertens}},
  \bibnamefont{and} \bibinfo{author}{\bibfnamefont{D.~D.} \bibnamefont{Sheka}},
  \emph{\bibinfo{title}{Effective anisotropy of thin nanomagnets: beyond the
  surface anisotropy approach}} (\bibinfo{year}{2007}{\natexlab{b}}),
  \eprint{arXiv:0705.1555}, \urlprefix\url{http://arxiv.org/abs/0705.1555}.

\bibitem[{OOM()}]{OOMMF}
\emph{\bibinfo{title}{The {O}bject {O}riented {M}icro{M}agnetic {F}ramework}},
  \bibinfo{note}{developed by M. J. Donahue and D. Porter mainly, from NIST. We
  used the 3D version of the 1.2$\alpha$2 release},
  \urlprefix\url{http://math.nist.gov/oommf/}.

\bibitem[{\citenamefont{Sheka}(2006)}]{Sheka06e}
\bibinfo{author}{\bibfnamefont{D.~D.} \bibnamefont{Sheka}},
  \bibinfo{journal}{Journal of Physics A: Mathematical and General}
  \textbf{\bibinfo{volume}{39}}, \bibinfo{pages}{15477} (\bibinfo{year}{2006}),
  \urlprefix\url{http://stacks.iop.org/0305-4470/39/15477}.

\bibitem[{\citenamefont{Choi et~al.}(2007)\citenamefont{Choi, Lee, Guslienko,
  and Kim}}]{Choi07}
\bibinfo{author}{\bibfnamefont{S.}~\bibnamefont{Choi}},
  \bibinfo{author}{\bibfnamefont{K.-S.} \bibnamefont{Lee}},
  \bibinfo{author}{\bibfnamefont{K.~Y.} \bibnamefont{Guslienko}},
  \bibnamefont{and} \bibinfo{author}{\bibfnamefont{S.-K.} \bibnamefont{Kim}},
  \bibinfo{journal}{Physical Review Letters} \textbf{\bibinfo{volume}{98}},
  \bibinfo{eid}{087205} (pages~\bibinfo{numpages}{4}) (\bibinfo{year}{2007}),
  \urlprefix\url{http://link.aps.org/abstract/PRL/v98/e087205}.

\bibitem[{\citenamefont{Hertel et~al.}(2007)\citenamefont{Hertel, Gliga,
  Fahnle, and Schneider}}]{Hertel07}
\bibinfo{author}{\bibfnamefont{R.}~\bibnamefont{Hertel}},
  \bibinfo{author}{\bibfnamefont{S.}~\bibnamefont{Gliga}},
  \bibinfo{author}{\bibfnamefont{M.}~\bibnamefont{Fahnle}}, \bibnamefont{and}
  \bibinfo{author}{\bibfnamefont{C.~M.} \bibnamefont{Schneider}},
  \bibinfo{journal}{Physical Review Letters} \textbf{\bibinfo{volume}{98}},
  \bibinfo{eid}{117201} (pages~\bibinfo{numpages}{4}) (\bibinfo{year}{2007}),
  \urlprefix\url{http://link.aps.org/abstract/PRL/v98/e117201}.

\bibitem[{\citenamefont{Waeyenberge et~al.}(2006)\citenamefont{Waeyenberge,
  Puzic, Stoll, Chou, Tyliszczak, Hertel, Fahnle, Bruckl, Rott, Reiss
  et~al.}}]{Waeyenberge06}
\bibinfo{author}{\bibfnamefont{V.~B.} \bibnamefont{Waeyenberge}},
  \bibinfo{author}{\bibfnamefont{A.}~\bibnamefont{Puzic}},
  \bibinfo{author}{\bibfnamefont{H.}~\bibnamefont{Stoll}},
  \bibinfo{author}{\bibfnamefont{K.~W.} \bibnamefont{Chou}},
  \bibinfo{author}{\bibfnamefont{T.}~\bibnamefont{Tyliszczak}},
  \bibinfo{author}{\bibfnamefont{R.}~\bibnamefont{Hertel}},
  \bibinfo{author}{\bibfnamefont{M.}~\bibnamefont{Fahnle}},
  \bibinfo{author}{\bibfnamefont{H.}~\bibnamefont{Bruckl}},
  \bibinfo{author}{\bibfnamefont{K.}~\bibnamefont{Rott}},
  \bibinfo{author}{\bibfnamefont{G.}~\bibnamefont{Reiss}},
  \bibnamefont{et~al.}, \bibinfo{journal}{Nature}
  \textbf{\bibinfo{volume}{444}}, \bibinfo{pages}{461} (\bibinfo{year}{2006}),
  ISSN \bibinfo{issn}{0028-0836},
  \urlprefix\url{http://dx.doi.org/10.1038/nature05240}.

\bibitem[{\citenamefont{Xiao et~al.}(2006)\citenamefont{Xiao, Rudge, Choi,
  Hong, and Donohoe}}]{Xiao06}
\bibinfo{author}{\bibfnamefont{Q.~F.} \bibnamefont{Xiao}},
  \bibinfo{author}{\bibfnamefont{J.}~\bibnamefont{Rudge}},
  \bibinfo{author}{\bibfnamefont{B.~C.} \bibnamefont{Choi}},
  \bibinfo{author}{\bibfnamefont{Y.~K.} \bibnamefont{Hong}}, \bibnamefont{and}
  \bibinfo{author}{\bibfnamefont{G.}~\bibnamefont{Donohoe}},
  \bibinfo{journal}{Applied Physics Letters} \textbf{\bibinfo{volume}{89}},
  \bibinfo{eid}{262507} (pages~\bibinfo{numpages}{3}) (\bibinfo{year}{2006}),
  \urlprefix\url{http://link.aip.org/link/?APL/89/262507/1}.

\bibitem[{\citenamefont{Lee et~al.}(2007)\citenamefont{Lee, Guslienko, Lee, and
  Kim}}]{Lee07}
\bibinfo{author}{\bibfnamefont{K.-S.} \bibnamefont{Lee}},
  \bibinfo{author}{\bibfnamefont{K.~Y.} \bibnamefont{Guslienko}},
  \bibinfo{author}{\bibfnamefont{J.-Y.} \bibnamefont{Lee}}, \bibnamefont{and}
  \bibinfo{author}{\bibfnamefont{S.-K.} \bibnamefont{Kim}},
  \emph{\bibinfo{title}{Ultrafast vortex-core reversal dynamics in
  ferromagnetic nanodots}} (\bibinfo{year}{2007}),
  \bibinfo{note}{cond-mat/0703538},
  \urlprefix\url{http://www.arxiv.org/abs/cond-mat/0703538}.

\bibitem[{\citenamefont{Tretiakov and Tchernyshyov}(2007)}]{Tretiakov07}
\bibinfo{author}{\bibfnamefont{O.~A.} \bibnamefont{Tretiakov}}
  \bibnamefont{and}
  \bibinfo{author}{\bibfnamefont{O.}~\bibnamefont{Tchernyshyov}},
  \bibinfo{journal}{Physical Review B (Condensed Matter and Materials Physics)}
  \textbf{\bibinfo{volume}{75}}, \bibinfo{eid}{012408}
  (pages~\bibinfo{numpages}{2}) (\bibinfo{year}{2007}),
  \urlprefix\url{http://link.aps.org/abstract/PRB/v75/e012408}.

\bibitem[{\citenamefont{Belavin and Polyakov}(1975)}]{Belavin75}
\bibinfo{author}{\bibfnamefont{A.~A.} \bibnamefont{Belavin}} \bibnamefont{and}
  \bibinfo{author}{\bibfnamefont{A.~M.} \bibnamefont{Polyakov}},
  \bibinfo{journal}{JETP Lett.} \textbf{\bibinfo{volume}{22}},
  \bibinfo{pages}{245} (\bibinfo{year}{1975}).

\bibitem[{\citenamefont{Hertel and Schneider}(2006)}]{Hertel06}
\bibinfo{author}{\bibfnamefont{R.}~\bibnamefont{Hertel}} \bibnamefont{and}
  \bibinfo{author}{\bibfnamefont{C.~M.} \bibnamefont{Schneider}},
  \bibinfo{journal}{Phys. Rev. Lett.} \textbf{\bibinfo{volume}{97}},
  \bibinfo{eid}{177202} (pages~\bibinfo{numpages}{4}) (\bibinfo{year}{2006}),
  \urlprefix\url{http://link.aps.org/abstract/PRL/v97/e177202}.

\bibitem[{\citenamefont{Hertel et~al.}(2006)\citenamefont{Hertel, Gliga,
  Fahnle, and Schneider}}]{Hertel06a}
\bibinfo{author}{\bibfnamefont{R.}~\bibnamefont{Hertel}},
  \bibinfo{author}{\bibfnamefont{S.}~\bibnamefont{Gliga}},
  \bibinfo{author}{\bibfnamefont{M.}~\bibnamefont{Fahnle}}, \bibnamefont{and}
  \bibinfo{author}{\bibfnamefont{C.~M.} \bibnamefont{Schneider}},
  \emph{\bibinfo{title}{Turning magnetic vortex cores upside down: Nanomagnetic
  toggle switching of vortex cores on the picosecond time scale}}
  (\bibinfo{year}{2006}),
  \urlprefix\url{http://arxiv.org/abs/cond-mat/0611668}.

\bibitem[{\citenamefont{Yamada et~al.}(2007)\citenamefont{Yamada, Kasai,
  Nakatani, Kobayashi, Kohno, Thiaville, and Ono}}]{Yamada07}
\bibinfo{author}{\bibfnamefont{K.}~\bibnamefont{Yamada}},
  \bibinfo{author}{\bibfnamefont{S.}~\bibnamefont{Kasai}},
  \bibinfo{author}{\bibfnamefont{Y.}~\bibnamefont{Nakatani}},
  \bibinfo{author}{\bibfnamefont{K.}~\bibnamefont{Kobayashi}},
  \bibinfo{author}{\bibfnamefont{H.}~\bibnamefont{Kohno}},
  \bibinfo{author}{\bibfnamefont{A.}~\bibnamefont{Thiaville}},
  \bibnamefont{and} \bibinfo{author}{\bibfnamefont{T.}~\bibnamefont{Ono}},
  \bibinfo{journal}{Nat Mater} \textbf{\bibinfo{volume}{6}},
  \bibinfo{pages}{270} (\bibinfo{year}{2007}), ISSN \bibinfo{issn}{1476-1122},
  \urlprefix\url{http://dx.doi.org/10.1038/nmat1867}.

\bibitem[{\citenamefont{Liu et~al.}(2007)\citenamefont{Liu, Gliga, Hertel, and
  Schneider}}]{Liu07}
\bibinfo{author}{\bibfnamefont{Y.}~\bibnamefont{Liu}},
  \bibinfo{author}{\bibfnamefont{S.}~\bibnamefont{Gliga}},
  \bibinfo{author}{\bibfnamefont{R.}~\bibnamefont{Hertel}}, \bibnamefont{and}
  \bibinfo{author}{\bibfnamefont{C.~M.} \bibnamefont{Schneider}},
  \emph{\bibinfo{title}{Current-induced magnetic vortex core switching in a
  permalloy nanodisk}} (\bibinfo{year}{2007}),
  \urlprefix\url{http://www.arXiv.org/cond-mat/0702048}.

\end{thebibliography}

\end{document}